\documentclass[10pt,a4paper,twocolumn]{article}
\usepackage{arxiv}

\usepackage[utf8]{inputenc} 
\usepackage[T1]{fontenc}    
\usepackage[colorlinks = true, 
            linkcolor = blue, 
            urlcolor = blue, 
            citecolor = blue, 
            anchorcolor = blue]{hyperref}       
\usepackage{url}            
\usepackage{booktabs}       
\usepackage{amsfonts}       
\usepackage{nicefrac}       
\usepackage{microtype}      
\usepackage{lipsum}
\usepackage{authblk}
\usepackage{amsmath} 
\usepackage{graphicx}
\usepackage{enumerate}
\usepackage{multicol}
\usepackage{subcaption}
\usepackage{bm}
\usepackage[super,sort&compress,comma]{natbib} 

\title{SMILES-X: \\
autonomous molecular compounds characterisation for small datasets without descriptors}

\author[1,*]{Guillaume Lambard}
\author[2,3]{Ekaterina Gracheva}

\affil[1]{Research and Services Division of Materials Data and Integrated System, Energy Materials Design Group, National Institute for Materials Science, 1-2-1 Sengen, Tsukuba, Ibaraki, 305-0047, Japan.}
\affil[2]{International Center for Materials Nanoarchitectonics, National Institute for Materials Science, 1-2-1 Sengen, Tsukuba, Ibaraki, 305-0047 Japan.}
\affil[3]{University of Tsukuba, 1-1-1 Tennodai, Tsukuba, Ibaraki, 305-8577 Japan}
\affil[*]{Corresponding author: LAMBARD.Guillaume@nims.go.jp}
\begin{document}

\twocolumn[
  \begin{@twocolumnfalse}
    \maketitle

\begin{abstract}
There is more and more evidence that machine learning can be successfully applied in materials science and related fields. However, datasets in these fields are often quite small ($\ll1000$ samples). It makes the most advanced machine learning techniques remain neglected, as they are considered to be applicable to big data only. Moreover, materials informatics methods often rely on human-engineered descriptors, that should be carefully chosen, or even created, to fit the physicochemical property that one intends to predict. In this article, we propose a new method that tackles both the issue of small datasets and the difficulty of task-specific descriptors development. The SMILES-X is an autonomous pipeline for molecular compounds characterisation based on a \{Embed-Encode-Attend-Predict\} neural architecture with a data-specific Bayesian hyper-parameters optimisation. The only input to the architecture — the SMILES strings — are de-canonicalised in order to efficiently augment the data. One of the key features of the architecture is the attention mechanism, which enables the interpretation of output predictions without extra computational cost. The SMILES-X shows new state-of-the-art results in the inference of aqueous solubility ($\overline{RMSE}_{test} \simeq 0.57 \pm 0.07$ mols/L), hydration free energy ($\overline{RMSE}_{test} \simeq 0.81 \pm 0.22$ kcal/mol, which is $\sim 24.5\%$ better than molecular dynamics simulations), and octanol/water distribution coefficient ($\overline{RMSE}_{test} \simeq 0.59 \pm 0.02$ for LogD at pH 7.4) of molecular compounds. The SMILES-X is intended to become an important asset in the toolkit of materials scientists and chemists. The source code for the SMILES-X is available at \href{https://github.com/GLambard/SMILES-X}{github.com/GLambard/SMILES-X}.


\end{abstract}
  \end{@twocolumnfalse}
  \vspace{0.3cm}

]

\section{Introduction}
In the fields of bio- and cheminformatics, machine learning (ML) algorithms combined with human-engineered molecular descriptors\cite{todeschini, desclist} have shown great potential in tasks of predicting physicochemical properties of molecular compounds. In practice, however, it is often necessary to run a blind scan through a large number of such combinations in order to find the most accurate inference model, which still may not lead to success. Most of the descriptors are task- or domain-specific, which makes their use impossible for more general problems, such as virtual screening, similarity searching, clustering and structure-activity modelling\cite{willett, cereto, mcgregor, li}. 

For these purposes molecular fingerprints have been developed. Fingerprint is a binary representation of a molecule: its structural or functional features are translated into a string of bits in the way to keep the fingerprint invariant to rotations, translations and property-preserving atomic permutations (see, e.g., extended circular fingerprints\cite{morgan}). Even though molecular fingerprints are known to be helpful to drugs discovery or compounds search among various databases, they may as well be detrimental to materials characterisation and design.
Therefore, while both descriptors and fingerprints may be beneficial, they come along with restrictions.

In fields like materials science it is common to have datasets with $\ll1000$ samples, which is considered to be too small for a direct deep learning application. Some research groups use neural architectures (NAs) for secondary tasks such as to build novel high-level features as non-linear combinations of molecular descriptors\cite{coley, mayr, ramsundar}. Others use NA to automatically learn features based on 2D/3D images\cite{goh, wallach}, molecular graphs\cite{duvenaud}, SMILES (simplified molecular input line entry system)\cite{bjerrum, hodas, kimber}, N-gram graphs\cite{liu} or a combination of mentioned inputs\cite{paul}, similar to computer vision (CV). Still, none of them intends to design an NA for property prediction on small datasets.
There are some works on transfer learning\cite{hutchinson, john, goh}, but the results vary greatly depending on the correlation between the tasks – which is often unknown \textit{a priori}. Moreover, most of the NAs used in the fields of CV or natural language processing (NLP) are trained on big data and impose architectures that do not fit small datasets.

Aside from the lack of data, another bottleneck on the way of using NAs in physics and chemistry is the lack of interpretability. A method for explaining neural networks has been recently proposed\cite{hodas}. It consists in training an additional neural network to generate a mask identifying the most important SMILES characters. Despite the respectable coherence in the interpretation of the chemical solubility, the explanation network is entirely correlated to its prediction network, which forces the training phase to be doubled for each dataset. Moreover, even though the explanation network allows to identify the groups having the highest weight in the property prediction, there is no evidence that the original prediction network has also learned the known chemistry concepts in order to make proper characterisation.

In this article we propose a method allowing to overpass the issues of data scarcity, descriptors engineering and the prediction interpretation ambiguity at the same time. The algorithm benefits from the natural ability of NAs to learn a suitable and task-specific representation of the data. It designs a simple yet effective NA dedicated to small datasets based on attention mechanism\cite{xu, chan, raffel}.  To achieve this, we borrowed the latest techniques from the CV and NLP fields to build an entirely autonomous system – the SMILES-X. To the best of our knowledge, this is the first time in materials science related fields when an NA is specifically designed to manage small datasets, and the first attempt to integrate a NLP-based attention mechanism for predicting physicochemical properties of molecular compounds. This mechanism allows to reduce the number of trainable parameters, and provides the interpretation of the results at no extra cost. The SMILES-X achieves the state-of-the-art results, predicting any physicochemical property given the molecule's SMILES\cite{daylight, blueobelisk} as the sole input.

The structure of the article is as follows. First, we describe the entire pipeline of the SMILES-X in Section \ref{sec:pipeline}. The SMILES augmentation and formatting are detailed in subsections \ref{par:augmentation}, \ref{par:tokenisation}, respectively, while the procedures of building the NA frame and its data-specific optimisation are presented in the subsection \ref{par:archsearch}. The subsection \ref{subsec:predictions} is dedicated to the performance of the SMILES-X based on three benchmark datasets for regression tasks from the MoleculeNet\cite{moleculenet}: ESOL\cite{esol}, FreeSolv\cite{freesolv} and Lipophilicity\cite{lipo}. There are three modes of interpretation of the results of the SMILES-X, which are discussed in the subsection \ref{subsec:interpret}. Finally, we conclude and discuss further possible improvements of the SMILES-X, as well as propose more potential target properties to be inferred using the algorithm in Section \ref{sec:conclusion}.

\section{The SMILES-X pipeline}
\label{sec:pipeline}
\begin{figure}[t]
\centering
\includegraphics[scale = 0.45]{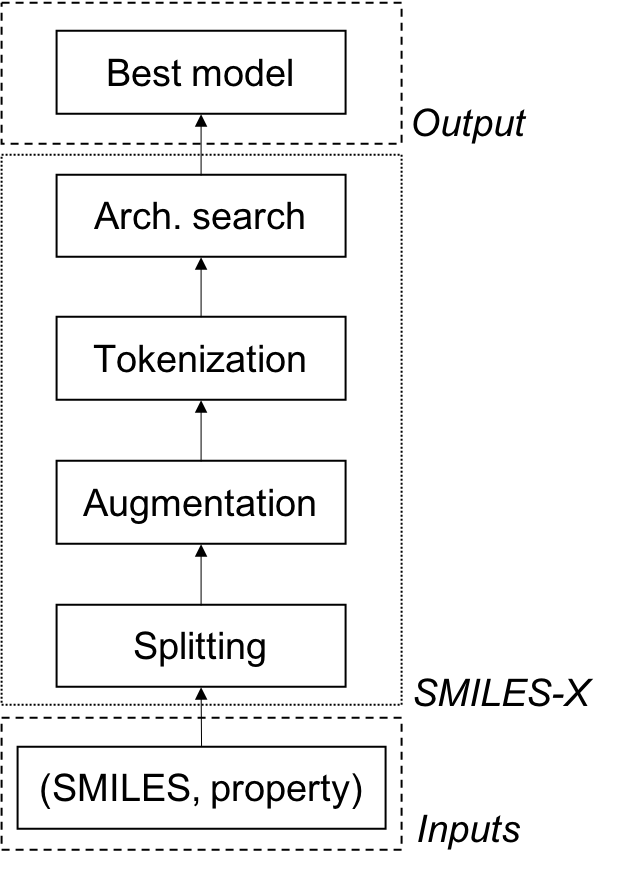}
\caption{The SMILES-X pipeline.}
\label{fig:fig1}
\end{figure}

The SMILES-X has been conceived to meet the following requirements: (i) to use the SMILES format as the only representation of a molecular compound; computable characteristics, such as the fingerprints or physical descriptors, are left out. (ii) Remove the SMILES canonicalization\cite{daylight} in order to exploit the full capacity of the molecular compound representation. (iii) The core architecture is simple enough to handle small datasets without sacrificing the prediction accuracy. (iv) Outcomes of the SMILES-X are interpretable.

Figure \ref{fig:fig1} is a sketch of the main steps within the SMILES-X pipeline. The primary input is a list of SMILES strings with corresponding property values. Then, a splitting into training, validation and test sets is performed via equiprobable sampling. The subsequent steps are detailed below.

\subsection{Augmentation}
\label{par:augmentation}
It has been shown in CV that data augmentation approaches such as flipping, rotation, scaling, cropping and other image transformations are effective to reduce the error rate on classification tasks and improve generalisation\cite{perez}. Here, we introduce a technique called SMILES augmentation, similar to Bjerrum\cite{bjerrum}. The first step consists in removing canonicalization\cite{daylight} of the SMILES. Canonicalization is the default procedure to standardise the SMILES across the databases, therefore removing it leads to an expanded number of SMILES individual representations. Then, augmentation is done by iterating over the following two steps: (i) Renumber the atoms of a given SMILES by rotation of their index. (ii) For each renumbering, reconstruct grammatically correct SMILES under the condition of conserving the initial molecule's isomerism and prohibiting kekulisation\cite{daylight, blueobelisk}. In the end, one obtains an expanded list of SMILES together with their corresponding property and cardinality $\rm n_{augm}(s_{i})$ (number of augmentations for a SMILES $\rm s_{i}$), if any. Duplicated SMILES are removed. The SMILES augmentation is individually performed after splitting into training, validation and test sets to avoid any information leakage. The procedure is performed using the RDKit library\cite{rdkit}.

\subsection{Tokenisation}
\label{par:tokenisation}
Tokenisation consists in dividing the SMILES into unique tokens, each token being a set of characters. The procedure of SMILES tokenisation is as follows\cite{daylight, blueobelisk}: (i) Aliphatic and aromatic organic atoms (B, C, N, O, S, P, F, Cl, Br, I, b, c, n, o, s, p), bounds, branches and rings (-, =, \#, \$, /, \textbackslash, ., (, ), \%digits, digit) are set as individual tokens. (ii) The characters between squared brackets, that may include inorganic and aromatic organic atoms, isotopes, chirality, hydrogen count, charges or class number, form a single token (brackets included, e.g., [NH4{\small +}]). (iii) Unlike the NLP analysis, the beginning token is not different from the termination one: both of them are represented by a whitespace, which is added at both ends of a tokenized SMILES. This is important to keep its reading direction invariant. Finally, a set of unique tokens is extracted to form the representative chemical vocabulary for a given dataset. To become an interpretable NA input, this vocabulary is then mapped into integers, and is conserved into memory for future usage. 

\subsection{Architecture search}
\label{par:archsearch}

\begin{figure}[t]
\centering
\includegraphics[scale = 0.45]{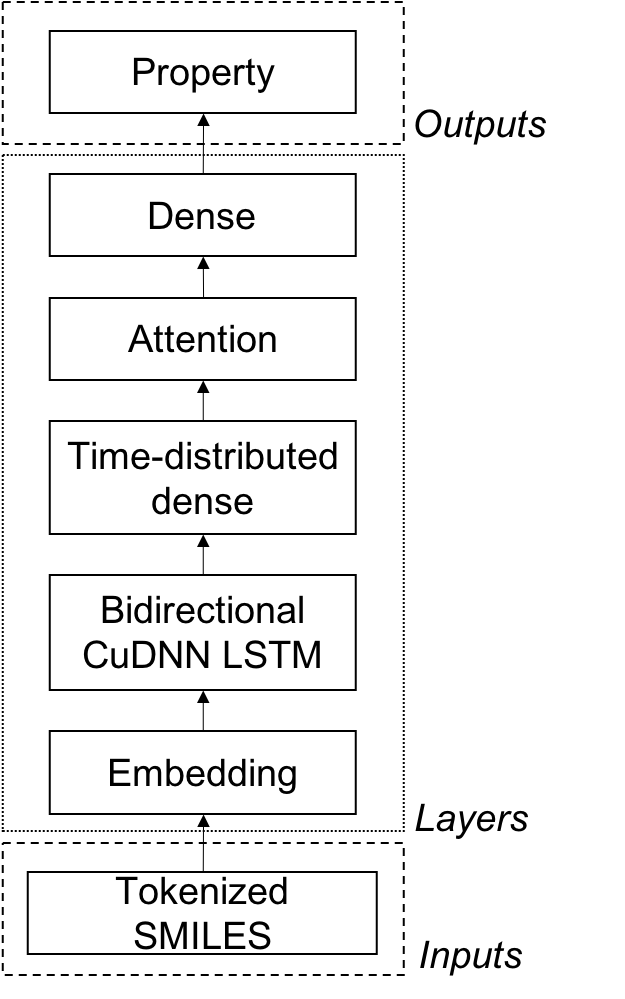}
\caption{Fixed skeleton of the neural architecture in the SMILES-X.}
\label{fig:skeleton}
\end{figure}

The neural architecture search has recently reached a new milestone in finding the optimal NA for a given task, by using, e.g., reinforcement learning techniques\cite{nas,enas} or evolutionary algorithms\cite{nasevol}. However, not only these techniques are computationally expensive but also they do not necessarily deal with the recurrent blocks. It has therefore been decided to fix the overall NA geometry (Figure \ref{fig:skeleton}) and search for the best set of the hyperparameters through the Bayesian optimisation\cite{frazier}. As it was mentioned earlier in Section \ref{sec:pipeline}, this geometry is NLP-oriented and treats SMILES strings as sentences in the chemical language; it has low complexity so as to be applicable to small datasets, and its outcomes are interpretable. Inspired by the hierarchical neural architecture\cite{yang}, which allows to get cutting edge results on document classification, we have built the SMILES-X frame based on a four-step formula: \{Embed, Encode, Attend, Predict\}\cite{honnibal}. 

\begin{enumerate}

\item \textbf{Embed}
The embedding layer\cite{gal} transforms the tokens, derived from the dataset's vocabulary in form of integers, into dense $\rm n_{embed}$-dimensional float vectors. Unlike arbitrary ordinal numbers, these vectors encapsulate the semantic meaning of tokens and their relations. This operation transforms SMILES into series of $\rm n_{embed} \times 1$ vectors, or $\rm n_{tokens} \times n_{embed}$ tensor, where $\rm n_{tokens}$ corresponds to the number of tokens in a tokenised SMILES string. 

\item \textbf{Encode}
The encoding phase is responsible for modifying the embedding, so that it captures the relationships between tokens in the context of the dataset. It consists of two neural layers: a bidirectional CuDNN long short-term memory (LSTM) layer\cite{lstm1, lstm2} is followed by a time-distributed fully connected one. The former consists of $\rm n_{LSTM}$ LSTM blocks and maps the input SMILES, represented now by a $\rm n_{tokens} \times n_{embed}$ tensor, into a context-aware $\rm n_{tokens} \times n_{LSTM}$ tensor. After training, each row of the tensor represents the meaning of a given token within the context of the rest of the SMILES string containing it. The bidirectionality forces the embedded SMILES to be sequentially passed forwards and backwards, conserving the invariance of their reading direction. The forward and backward encodings of a SMILES are then concatenated, resulting in a $\rm n_{tokens} \times 2n_{LSTM}$ output tensor. The time-distributed dense layer is then applied to each of $\rm n_{tokens}$ tokens. This allows to capture the relationships between tokens in greater detail, or in other words to deepen the LSTM layer (similar to the effect of adding an extra dense layer to a vanilla neural network). Given that the number of hidden units in this layer is $\rm n_{dense}$, the output after encoding is a $\rm n_{\rm tokens} \times n_{dense}$ tensor. It should be noted that we specifically use CuDNN LSTM\cite{cudnn} blocks for efficient optimization and training phases on GPU from NVIDIA Corporation. Without the CuDNN version of LSTM, the speed of training would drop by a factor of $\sim 10$, making the optimisation phase intractable.

\item \textbf{Attend}
The attention layer detects the salient tokens, compressing tensor $\rm H \in \mathbb{R}^{n_{tokens} \times n_{dense}}$ into an $\rm n_{dense}$ vector c with minimum information loss\cite{raffel}:

\begin{flalign}
e &= tanh(\rm H \cdot W_{a} + b_{a}) \, , \nonumber \\
\alpha &= \frac{\exp(e)}{\sum_{i=1}^{\rm n_{tokens}} \exp(e_{i})} \, , \nonumber \\
c &= \rm H^{T} \cdot \alpha \, ,
\label{eq:att}
\end{flalign}

where $W_{a} \in \mathbb{R}^{\rm n_{dense} \times 1}$ and $b_{a}\in \mathbb{R}^{\rm n_{tokens} \times 1}$ are trainable parameters, $\alpha \in \mathbb{R}^{\rm n_{tokens} \times 1}$ is the attention vector and $c \in \mathbb{R}^{\rm n_{dense} \times 1}$ is the output. Thus, the attention layer performs two important tasks at once: (1) it collapses the representation $\rm H$ of a variable length chain of tokens into a fixed length vector c by applying a weighted sum over the tokens to fit the final property best, with (2) the weights in $\alpha$ which represent the importance of each token towards the final property prediction, bringing to a straightforward interpretation. Therefore, the attention layer has two modes, one returning the output vector c, and the other – the attention vector $\alpha$ (see Section \ref{sec:results}). The two modes are switchable at will without extra computational cost. 

\item \textbf{Predict}
The final NA layer transforms the attention layer output c into a single property value $\rm Prop(s_{i})$ by a simple linear operation:

\begin{equation}
\rm Prop(s_{i}) = W_{p}^{T} \cdot c + b_{p}\, ,
\label{eq:poutput}
\end{equation} 

The interpretation from $\alpha$ in Equation~\ref{eq:att} and the prediction are thus linearly connected and are accessible without any additional treatments on the input data or NA, unlike the pipelines in other works\cite{montavon, schutt, hodas}.
\end{enumerate}

It should be noted that all the above tensors or vectors have one additional dimension, $\rm n_{SMILES}$, omitted for the sake of simplicity. This dimension corresponds to the batch size of a single iteration passed to the network, i.e. the maximum number of SMILES that it processes at once. All of the steps above are implemented in Keras API\cite{keras} and Tensorflow\cite{tensorflow} with GPU support.

\section{Results \& discussion}
\label{sec:results}

To evaluate the regression performance of the SMILES-X, it was chosen to test it on three benchmark physical chemistry datasets issued from the MoleculeNet\cite{moleculenet}. These datasets are considered as small, with less than 5000 compound-property pairs, and therefore present a challenge to machine learning models. The ESOL\cite{esol} dataset contains the logarithmic aqueous solubility (mols/L) for 1128 organic small molecules; the FreeSolv\cite{freesolv} consists of the calculated and experimental hydration free energies (kcal/mol) for 642 small neutral molecules in water; and the Lipophilicity\cite{lipo} stores the experimental data on octanol/water distribution coefficient (logD at pH 7.4) for 4200 compounds. 

In present report the splitting ratio for training/validation/test is set to 0.8/0.1/0.1.  Following the procedure from MoleculeNet\cite{moleculenet}, we performed 8 splits, each time using new seed for the Monte-Carlo sampling. The seeds have been fixed for the sake of reproducibility. We use the averaged RMSE over the 8 test sets as the comparison metric of performance.

The optimal model architecture is determined via Bayesian optimisation individually for each split. We used the python library GPyOpt\cite{gpyopt} for this purpose. The search bounds are as follows: ($\rm n_{embed},\ n_{LSTM}, n_{dense}\ and\ n_{SMILES}) \in \{8, 16, 32, 64, 128, 512, 1024\},\ \gamma \in [2; 4]$ with a step of 0.1, where $\gamma$ is related to the optimiser learning rate as $\rm lr \equiv 10^{-\gamma}$, making a total of 50421 configurations. For the Lipophilicity dataset, $\rm n_{SMILES}$ and learning rate are fixed to 1024 and $10^{-3}$, respectively, leaving 343 potential architectures to search among. First, 25 architectures are randomly sampled and trained. Then, a maximum of 25 architectures are proposed via the expected improvement acquisition function\cite{jones}. Each of the architectures are sequentially trained for 30 epochs for ESOL and FreeSolv, and 10 for the Lipophilicity set (these values have been chosen based on the speed/efficiency ratio). The best proposed architecture is finally trained using a standard Adam optimiser\cite{adam} with checkpoint and early stopping. The early stopping is configured to stop the training if the validation loss is not improving for 50 consecutive epochs, and a checkpoint saves the parameters of the model with the minimal validation loss. The maximum number of epochs is set to 300, but because of the early stopping condition this value has never been reached. Depending on whether the SMILES augmentation is requested or not, the code needs from 1 to 4 GPUs running in parallel.

\subsection{Predictions}
\label{subsec:predictions}

\begin{table*}
\small
\caption{Comparison of physicochemical properties predictions from the SMILES-X (Can,
Augm) to the best performances in MoleculeNet\cite{moleculenet} on the ESOL\cite{esol}, FreeSolv\cite{freesolv} and Lipophilicity\cite{lipo} datasets, and to molecular dynamics calculations\cite{freesolv} for the FreeSolv dataset only}
  \centering
  \begin{tabular*}{\textwidth}{@{\extracolsep{\fill}}llll}
    \hline
     & \multicolumn{3}{c}{Datasets} \\
    Method & ESOL & FreeSolv & Lipophilicity \\
    \hline
    MoleculeNet\cite{moleculenet}  & $0.58 \pm 0.03$ & $1.15 \pm 0.12$ & $0.65 \pm 0.04$ \\
    Molecular dynamics\cite{freesolv} & --- & $1.51 \pm 0.07$ & --- \\
    SMILES-X (Can) & $0.70 \pm 0.05$ &  $1.14 \pm 0.17$ & $0.68 \pm 0.05$ \\
    SMILES-X (Augm) & $\pmb{0.57 \pm 0.07}$ & $\pmb{0.81 \pm 0.22}$ & $\pmb{0.60 \pm 0.04}$ \\
    \hline
  \end{tabular*}
  \label{tab:results}
\end{table*}

\begin{figure}[t]
\centering
\includegraphics[scale=0.55]{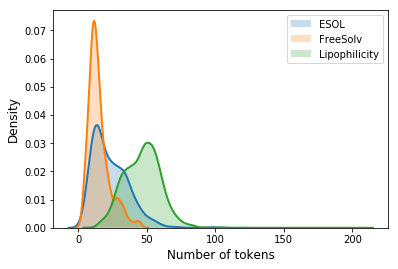}
\caption{Number of tokens per SMILES for the datasets ESOL~\cite{esol}, FreeSolv~\cite{freesolv}, and Lipophilicity~\cite{lipo}.}
\label{fig:stats}
\end{figure}

We compare the performance of SMILES-X against the best-to-date results from MoleculeNet\cite{moleculenet}, and for the FreeSolv additionally to the calculations based on the molecular dynamics simulations\cite{freesolv} (Table \ref{tab:results}). The results in MoleculeNet\cite{moleculenet} are reported for the molecular graph-based models that achieved the best results on a given dataset: concretely, a message passing neural network\cite{gilmer} for the ESOL and FreeSolv datasets, and a graph convolutional model\cite{gc} for the Lipophilicity dataset. Bayesian optimisation is also used there for the layers size, batch size and learning rate. We include both the results on canonicalised SMILES (Can) and on SMILES that have been augmented (Augm) (see Section \ref{par:archsearch}). When a SMILES string $\rm s_{i}$ is augmented to $\rm n_{augm}$ strings, its predicted property value is averaged over $\rm n_{augm}$ predictions. Table \ref{tab:results} shows that the SMILES-X reaches the best results for the FreeSolv and Lipophilicity datasets, improving the prediction accuracy by ~30\% and ~9\%, respectively, while having a comparable performance on the ESOL data. It is unclear why our algorithm fails to improve on the ESOL data. We thought that the number of tokens per SMILES may be the culprit. However, Figure \ref{fig:stats} shows that this is not the case. Note that even using the standard canonicalised SMILES strings, the property can be predicted quite well without employing any chemical knowledge (i.e., using no descriptors). Interestingly, machine learning allows to achieve a better accuracy than the molecular dynamics simulations.

There are the three main reasons that we think permitted SMILES-X to achieve these results:
\begin{enumerate}[i.]

\item The success is mainly attributed to the attention layer, that shows similar improvements in document classification tasks\cite{yang}. Comparing our performance to a similar NA without an attention layer\cite{hodas}, we see some 32.5\% improvement on accuracy.

\item Bayesian optimisation is a valuable tool that allows to efficiently find the best hyper-parameters in a short time.

\item It is obvious that SMILES augmentation shows great improvement (Can versus Augm in Table~\ref{tab:results}), and was necessary to achieve the best current results. Also, one can note that a graph-based NA would not allow such data augmentation.

\end{enumerate}

\begin{figure}[t]
\centering
\includegraphics[scale=0.4]{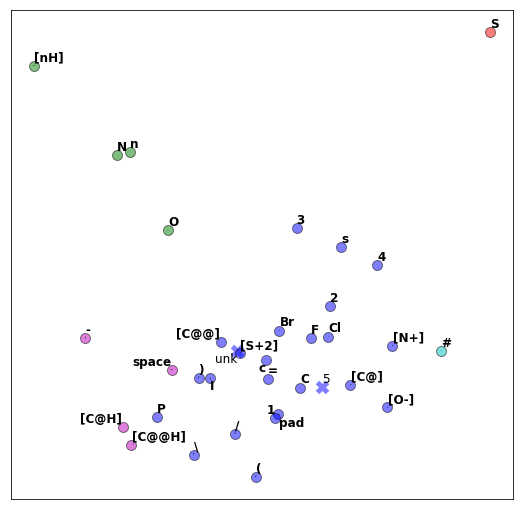}
\caption{Visualisation of a representation of SMILES tokens from the embedding layer for the FreeSolv~\cite{freesolv} dataset.}
\label{fig:embedrepr}
\end{figure}

\subsection{Interpretability}
\label{subsec:interpret}

\begin{figure}[t]
\centering

\begin{subfigure}[b]{\columnwidth}
\centering
\includegraphics[scale=0.25]{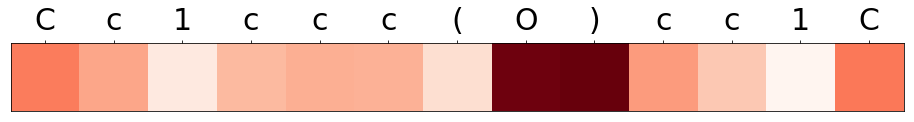}
\caption{}
\label{fig:interpn1d}
\end{subfigure}

\begin{subfigure}[b]{\columnwidth}
\centering
\includegraphics[scale=0.3]{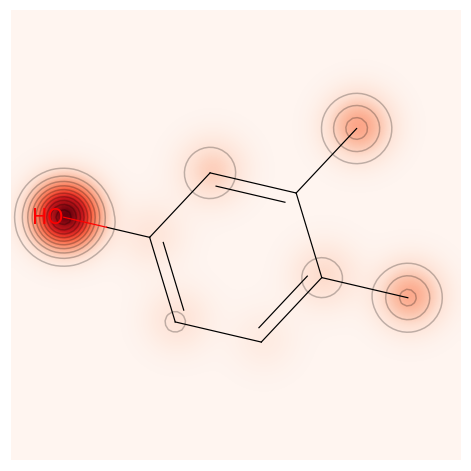}
\caption{}
\label{fig:interpn2d}
\end{subfigure}

\begin{subfigure}[b]{\columnwidth}
\centering
\includegraphics[scale=0.24]{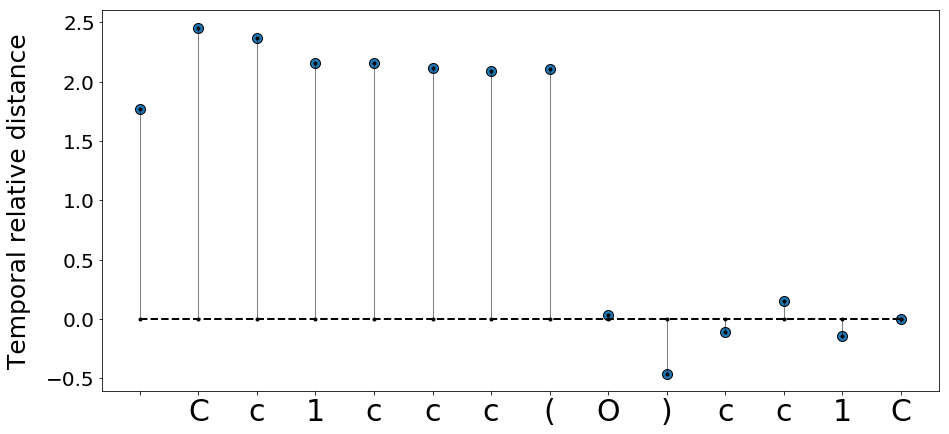}
\caption{}
\label{fig:interpntemp}
\end{subfigure}
\caption{Visualisation of the importance of each token within the SMILES towards the final prediction of the property of interest. The illustration is done on the structure Cc1ccc(O)cc1C from the FreeSolv dataset, with hydration free energy as the corresponding property. The 1D (a) and 2D (b) attention maps show the projections of the attention vector $\alpha$ on the SMILES string and molecular graph, respectively. The redder and darker the colour is, the stronger is the attention on a given token. The temporal relative distance $\rm \mathcal{T}_{dist}$ is shown in (c). The closer to zero is the distance value, the closer is the temporary prediction on the SMILES fragment to the whole SMILES prediction.}
\label{fig:interpns}
\end{figure}

As it was mentioned before, one of the great advantages of our method is its interpretability. The Figure 4 shows an example of the trained token embeddings. We used a principal component analysis (PCA\cite{pca1, pca2}) to reduce dimensionality from $\rm n_{embed} = 1024$ down to two, for the purpose of visualisation. The tokens that are not included in the training set, and are therefore randomly assigned, are represented by a cross.  One can see that halogens {Br, F, Cl} are located near each other. Other distinguishable sets are, for example, $\rm \left\{ [C@@],[S+2],c,C,[C@] \right\}$ and $\rm \left\{ n,N \right\}$, that have the same valence and bonds type within the group. The model also puts $\rm \left\{ [N+],[O-] \right\}$ close to each other, which reveals their regular coexistence in compounds within the FreeSolv data. Some other tokens placements, however, are not obvious to chemically qualify. In any case, the principle aim of clustering is to smooth out the chemical relations; it serves as a trainable look-up table for the further context-aware processing of tokens. We should not, thus, expect too great a degree of interpretability at this step. Representation of the individual tokens out of their chemical context is not the objective of the SMILES-X. 

Instead, we are interested in the interpretation of the network property prediction. With the SMILES-X, we are able to visualise the importance of each single token towards the final prediction of the property of interest (Figure 5).

There are three ways of visualisation available: (a) a 1D map built from the attention vector $\alpha$ (see Equation \ref{eq:att}) juxtaposed with the SMILES string, (b) a similar 2D version for the molecular graph and (c) temporal relative distance $\rm \mathcal{T}_{dist}$ to the predicted property. For the first two, the redder and darker the colour is the stronger is the attention on a given token. 

$\rm \mathcal{T}_{dist}(n)$ shows the evolution of the prediction for the SMILES while reading it token by token from left to right. It is inspired by Lanchantin\cite{lanchantin} and defined as:
\begin{equation}
\rm \mathcal{T}_{dist}(n) = \frac{Prop(n)-Prop(n_{tokens})}{|Prop(n_{tokens})|} \, ,
\label{eq:tempdist}
\end{equation}

where Prop(n) is the property predicted value based on the first n tokens of the SMILES for
$\rm n \in [1,...,n_{tokens}]$. Note that it converges to the final prediction $\rm Prop(n_{tokens}) \equiv Prop(s_{i})$ (prediction based on the entire SMILES). This also allows to judge as to how much a token influences the property of a compound. In this example, the prediction based on fragment 'Cc1ccc(O' is almost identical to the final prediction on the whole structure.

For the compound that we used as an example, the oxygen atom ('O') is considered to be the most influential element of the molecule for the hydration free energy prediction, which reflects chemical reality.

\section{Conclusions}
\label{sec:conclusion}
A new neural architecture for the chemical compounds characterisation, the SMILES-X, has been developed. In this article, we have presented the pipeline and performance of the SMILES-X. We demonstrate its aptitude to provide state-of-the-art results on the inference of several physicochemical properties, concretely the logarithmic aqueous solubility ($\overline{RMSE}_{test} \simeq 0.57 \pm 0.07$ mols/L), hydration free energy ($\overline{RMSE}_{test} \simeq 0.81 \pm 0.22$ kcal/mol) and octanol/water distribution coefficient ($\overline{RMSE}_{test} \simeq 0.60 \pm 0.04$ for LogD at pH 7.4). These results prove that it is now possible to successfully predict a physicochemical property employing no chemical intuition, even with a small dataset at hand.
The success of the SMILES-X rides on three key factors: (i) The Embed-Encode-Attend-Predict architecture, that allows to simplify the whole architecture thanks to the attention mechanism (i.e., to have less trainable parameters), and therefore reduces the risk of over-fitting. (ii) The Bayesian optimisation of the neural network's hyper-parameters allows to achieve close-to-optimal representation of the molecular compounds, per task and dataset. (iii) The use of SMILES strings as a sole input representation of chemical compounds allows efficient data augmentation.
 
Thanks to the attention mechanism, the SMILES-X comes with three modes of interpretation of the inference outcomes. This provides the end-user with the insights on which fragments of the chemical structure have the highest (or the lowest) influence on the property of interest. This kind of artificial intuition is a valuable asset not only for the tasks of characterisation and design of novel compounds, but also to re-purpose already-known materials. 

As for the future improvement on the SMILES-X, we plan to use BERT-like\cite{bert} NA's skeleton for the sake of reducing the accuracy gap existing between the ESOL, FreeSolv and Lipophilicity datasets studied here. The LSTM blocks are known to have memory problems with very distant dependencies within long sentences, and an architecture that is entirely based on the attention mechanism, i.e. free from LSTM blocks, like BERT, may overcome this weakness. Another way to improve the inference accuracy may be via informative sampling\cite{infsampl}.

In our forthcoming article we will address the tasks of classification, still using the MoleculeNet's datasets\cite{moleculenet}. That means that the SMILES-X will be modified in order to handle single-to-many, many-to-many and many-to-single classification tasks.

\section*{Conflicts of interest}
There are no conflicts to declare.

\section*{Acknowledgements}
The authors gratefully acknowledge the NVIDIA Corporation for the Titan V and Titan Xp GPUs, without which this research would not have been possible.

\bibliographystyle{iopart-num}
\bibliography{SMILESX_20190704} 

\end{document}